\documentclass{iopjournal}

\usepackage{siunitx}
\usepackage{amsmath}
\DeclareRobustCommand{\rev}[1]{\textbf{#1}}

\newif\ifanon      \anonfalse
\newif\ifmarked    \markedfalse

\ifmarked
  \DeclareRobustCommand{\rev}[1]{\textbf{#1}}
\else
  \DeclareRobustCommand{\rev}[1]{#1}
\fi

\begin{document}

\articletype{Article type} 

\title{Infrasound Newtonian Noise Estimation at the Einstein Telescope Candidate Site Sos Enattos}

\ifanon
  \author{Anonymous}
\else
\author{Wathela Alhassan$^{1,3}$\orcid{0000-0002-8266-3005}, D. Gondek-Rosinska$^3$\orcid{0000-0000-0000-0000}, M. Suchenek$^{1,3}$\orcid{0000-0003-1865-2894}, E. Fenyvesi$^2$\orcid{0000-0003-2777-3719}, and T. Bulik$^3$\orcid{0000-0003-2045-4803}}

\affil{$^1$Department of Physics, College of Arts and Sciences, American University of Sharjah, PO Box 26666, Sharjah, United Arab Emirates}

\affil{$^2$HUN-REN Wigner Research Centre for Physics, Konkoly-Thege Miklós út 29-33., 1121 Budapest, Hungary}

\affil{$^3$Astronomical Observatory, University of Warsaw, Aleje Ujazdowskie 4, 00-478 Warsaw, Poland}

\affil{$^*$Author to whom any correspondence should be addressed.}

\email{wathelahamed@gmail.com}
\fi


\begin{abstract}
We investigate the seasonal variability of atmospheric infrasound and its contribution to Newtonian noise (NN) at the Sos Enattos site, a leading candidate location for the Einstein Telescope (ET). Infrasound data recorded at three stations, SOE0 (surface), SOE1 ($-84$~m) and SOE3 ($-160$~m), are analysed over multiple seasons to characterise both the temporal variability and the depth dependence of the acoustic field. At the surface, the median amplitude spectral density (ASD) at 1~Hz shows no consistent seasonal modulation, and its variability is driven primarily by wind; the apparent dependence on sea state largely reflects its correlation with wind speed. Underground, where local wind noise is suppressed, a winter maximum of 4--7~dB above the quietest season is observed in both years of data. Using the measured sound-pressure spectra, we estimate the corresponding NN contribution for \rev{an underground cavern with the dimensions foreseen for ET}, assuming an isotropic field of acoustic plane waves. \rev{The surface pressure field, acting through the atmosphere above a test mass at 200~m depth, exceeds the ET-D design sensitivity in median conditions between 1.6 and 7.1~Hz, by up to a factor of 4.7. Inside the cavern, the median NN remains a factor of 1.4 below ET-D with the pressure measured at SOE1, but exceeds it between 2.5 and 8.7~Hz, by up to a factor of 3.5, with the pressure measured at SOE3.} The attenuation is not monotonic with depth: the deeper station SOE3 shows higher levels than SOE1 above a few hertz, pointing to local, anthropogenic sources in the underground infrastructure. These results show that underground placement is essential \rev{but not sufficient on its own}, and that the acoustic emission of \rev{the underground infrastructure and} future laboratory equipment will require careful evaluation and mitigation to keep atmospheric NN below the ET design sensitivity.
\end{abstract}
\section{Introduction}
The Einstein Telescope (ET) \cite{Punturo2010,Abernathy2011,Maggiore2020} is a proposed third-generation ground-based gravitational-wave observatory designed to achieve an order-of-magnitude improvement in sensitivity compared to current detectors such as Advanced LIGO \cite{Aasi2015} and Advanced Virgo \cite{Acernese2015}, which opened the field of gravitational-wave astronomy with the first direct detection of a binary black hole merger \cite{Abbott2016}. Together with the proposed Cosmic Explorer observatory \cite{Reitze2019}, ET will define the next generation of gravitational-wave facilities. A key design goal of ET is to extend the observational band toward low frequencies (from 1 to 10 Hz), where numerous astrophysical sources, including massive black hole mergers and early inspiral phases of compact binaries, can be observed \cite{Maggiore2020,Branchesi2023}. However, in this frequency range, the detector sensitivity, as expressed for example by the ET-D design curve \cite{Hild2011}, is fundamentally limited by environmental noise sources.
 
Among these, Newtonian noise (NN), also referred to as gravity-gradient noise, represents a major challenge for terrestrial gravitational-wave detectors. NN arises from time-dependent fluctuations of the local gravitational field caused by moving masses in the environment, such as seismic waves, atmospheric density perturbations, and human activity \cite{Saulson1984,Beccaria1998,Harms2015,Harms2019}. Unlike other noise sources, NN cannot be shielded by conventional isolation systems, as it directly couples to the test masses via gravitational interaction. Therefore, its mitigation relies primarily on careful site selection, underground installation \cite{Beker2011}, passive suppression strategies such as topographic shaping and recess structures \cite{HarmsHild2014}, and advanced subtraction techniques based on optimized sensor arrays \cite{Driggers2012,Coughlin2016,Badaracco2019}. A potentially significant, yet still insufficiently constrained, component of NN is associated with atmospheric infrasound. Infrasound waves, typically defined as acoustic waves below 20 Hz, constitute a persistent component of the ambient atmospheric noise background \cite{Bowman2005} and produce fluctuations in air density that generate time-varying gravitational forces acting on the detector test masses. This effect becomes particularly relevant in the under 10 Hz frequency range, which is critical for ET sensitivity \cite{Fiorucci2018,Brundu2022}. The magnitude of this contribution depends on several factors, including atmospheric conditions, local meteorology, and site-specific properties such as topography and underground depth. The identification of suitable detector sites with exceptionally low environmental noise is therefore a central task of the Einstein Telescope Collaboration \cite{Amann2020,DiGiovanni2025}, and comparable characterization campaigns are being carried out at the other ET candidate site in the Euregio Meuse--Rhine \cite{Bader2022}. The Sos Enattos mining area in Sardinia, Italy, has emerged as one of the most promising candidate locations due to its low seismicity, low anthropogenic noise, and favorable geological conditions \cite{Naticchioni2014,Naticchioni2020,DiGiovanni2021,Allocca2021,Saccorotti2023}. Long-term monitoring at the site has further quantified the temporal variability of the ambient seismic field \cite{DiGiovanni2023} and the stability of the surrounding ground \cite{Dessi2023}. In particular, underground environments offer significant advantages, as they provide natural shielding from surface disturbances and reduce the coupling of atmospheric and seismic noise to the detector.
 
Previous studies have investigated various contributions to Newtonian noise, including seismic, atmospheric, and thermal effects, and have demonstrated that suppression by several orders of magnitude will be required for next-generation detectors \cite{Harms2019,Harms2022lower}. Early work by \cite{Hughes1998} provided the first quantitative estimates of gravity-gradient noise in terrestrial interferometers, including contributions from both seismic and atmospheric sources. Subsequent analyses \cite{Creighton2008,Fiorucci2018,Brundu2022} have shown that atmospheric infrasound, as well as temperature-driven density fluctuations and turbulence, can generate measurable gravitational perturbations. These studies also highlighted the importance of underground siting and infrastructure design in mitigating such effects, while related work has examined the impact of correlated seismic and Newtonian noise on the triangular ET configuration \cite{Janssens2022}. Despite all these advances, the seasonal variability of infrasound and its depth dependence at candidate ET sites remain insufficiently characterized. In particular, long-term measurements that simultaneously probe surface and underground environments are still limited, and their implications for Newtonian noise estimates require further investigation.
 
In this work, we present the first comprehensive seasonal and depth-resolved analysis of infrasound measurements at the Sos Enattos site, based on data collected from three stations located at the surface (SOE0) and at underground depths of $-84~\mathrm{m}$ (SOE1) and $-160~\mathrm{m}$ (SOE3). Using these observations, we quantify the variability of infrasound noise as a function of season and environmental conditions, and we estimate the corresponding contribution to atmospheric Newtonian noise. Our results demonstrate a strong suppression of infrasound-induced NN with increasing depth and show that, even under unfavorable conditions, the atmospheric contribution remains well below the ET-D design sensitivity curve \cite{Hild2011} in the 1 to 10 Hz frequency range. These findings confirm that atmospheric infrasound does not pose a limiting factor for underground gravitational-wave detectors at Sos Enattos and further support its suitability as a candidate site for the Einstein Telescope.

\section{Materials and Methods}
Infrasound measurements at the Sos Enattos site were performed using condenser microphones deployed at multiple depths, enabling long-term, broadband monitoring of sound-pressure fluctuations, in line with established practice in low-frequency acoustic monitoring \cite{Ponceau2010, Marty2019}. The primary objective of the experiment was to characterise the infrasound conditions within the Sos Enattos mine, in the context of gravitational-wave detector studies \cite{Harms2015, Fiorucci2018}.

Each of the three measurement stations (SOE0, SOE1 and SOE3) was equipped with a single GRAS 47AC microphone. The microphone covers a frequency range from \SI{0.09}{\hertz} up to \SI{15}{\kilo\hertz}, with a sensitivity of \SI{8}{\milli\volt\per\pascal} and a dynamic range extending from a thermal noise floor of 20~dB(A) (A-weighted, as specified by the manufacturer) up to a maximum sound pressure level of 148~dB (unweighted), ensuring reliable detection of both weak and strong acoustic signals. The sensor provides a constant-current power (CCP) output, ensuring long-term stability and suitability for underground deployment.

At the underground stations SOE1 and SOE3, the microphone was installed in a dedicated cavity of $\SI{4}{\metre}\times\SI{4}{\metre}\times\SI{3}{\metre}$ (length $\times$ width $\times$ height), separated from the main tunnel by a thin wall made of polystyrene foam, partially covered with metal sheets and polyurethane foam. The purpose of the cavity is to reduce the direct influence of air movements on the microphone. A microphone placed directly in the tunnel is exposed to airflow induced by external wind and by the mine ventilation, which introduces additional pressure fluctuations and raises the noise floor. This was observed with additional microphones that were installed in the tunnel, outside the cavities, during the measurement campaign and are not used in this work. Their spectra showed the same characteristic spectral peaks at the same frequencies as the microphones inside the cavities, but a higher background between these peaks, which reduced the visibility of lower-amplitude features across the measured frequency range. The cavity therefore minimises the contribution of local airflow while preserving the acoustic fluctuations of interest.

The recorded data include contributions from both natural and anthropogenic sources relevant to Newtonian noise studies \cite{Bowman2005, Saulson1984}. In addition, co-located meteorological measurements and sea wave height data were used to investigate the influence of atmospheric conditions and marine activity on infrasound variability, the latter being a well-established driver of low-frequency environmental noise \cite{LonguetHiggins1950, Ardhuin2015, Naticchioni2020}.

\subsection{Site Description}

\begin{figure}[h!]
    \centering
    \includegraphics[width=\textwidth, height=8cm, keepaspectratio]{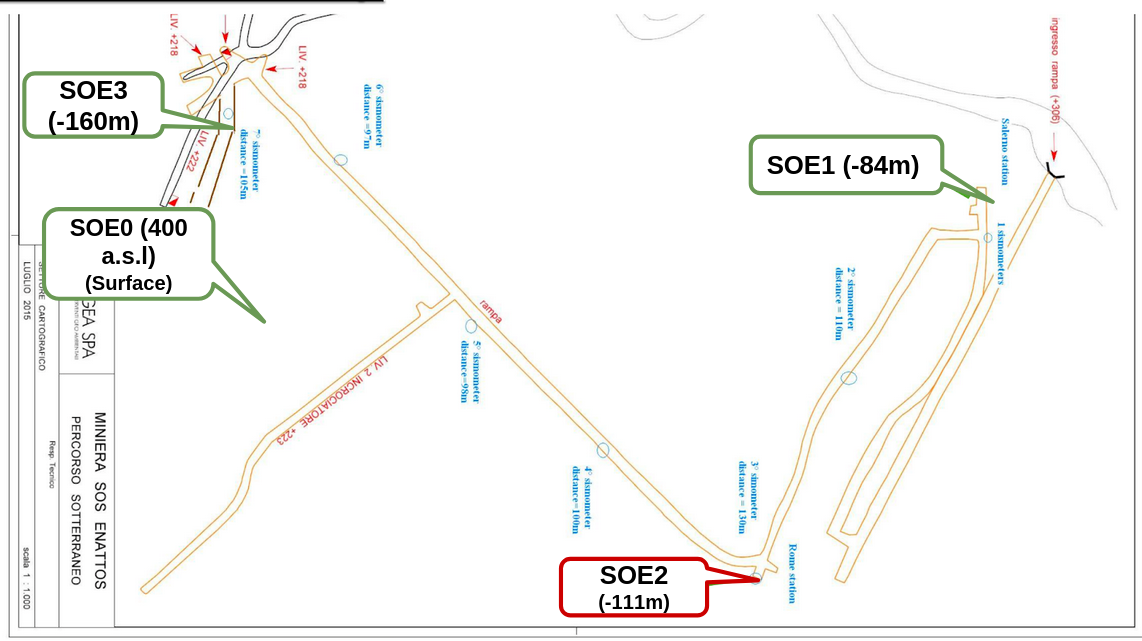}
    \caption{Layout of the Sos Enattos site showing the locations of the stations SOE0, SOE1, SOE2 and SOE3 at different depths. SOE0, SOE1 and SOE3 are each equipped with one GRAS 47AC microphone; SOE2, shown in red, has no microphone installed.}
    \label{map}
\end{figure}

The measurement campaign was conducted at three stations located at different depths within the Sos Enattos mine: SOE0, the surface station at \SI{400}{\metre} above sea level; SOE1, the first underground station, at a depth of \SI{84}{\metre}; and SOE3, the second underground station, at a depth of \SI{160}{\metre}. Each station was equipped with one GRAS 47AC microphone. A further underground site, SOE2, is shown for reference but was not instrumented with a microphone. The installation layout is shown in Figure~\ref{map}.

The surface station SOE0 was installed in a small roofless building of about $\SI{5}{\metre}\times\SI{5}{\metre}$ without windows, previously used for storing explosives for the mining operations, and powered by solar panels installed outside the building. To protect the microphone from direct exposure to wind and rain, it was placed inside a dedicated protective enclosure of about $\SI{40}{\centi\metre}\times\SI{40}{\centi\metre}$. This protection is particularly important because the microphone diaphragm is metallic and would otherwise be directly exposed to environmental conditions. The enclosure reduces the influence of wind impinging on the microphone capsule and of precipitation, while remaining acoustically transparent at infrasonic frequencies, whose wavelengths are much larger than the enclosure.

\begin{figure}
    \centering
    \includegraphics[width=0.9\textwidth, height=8cm, keepaspectratio]{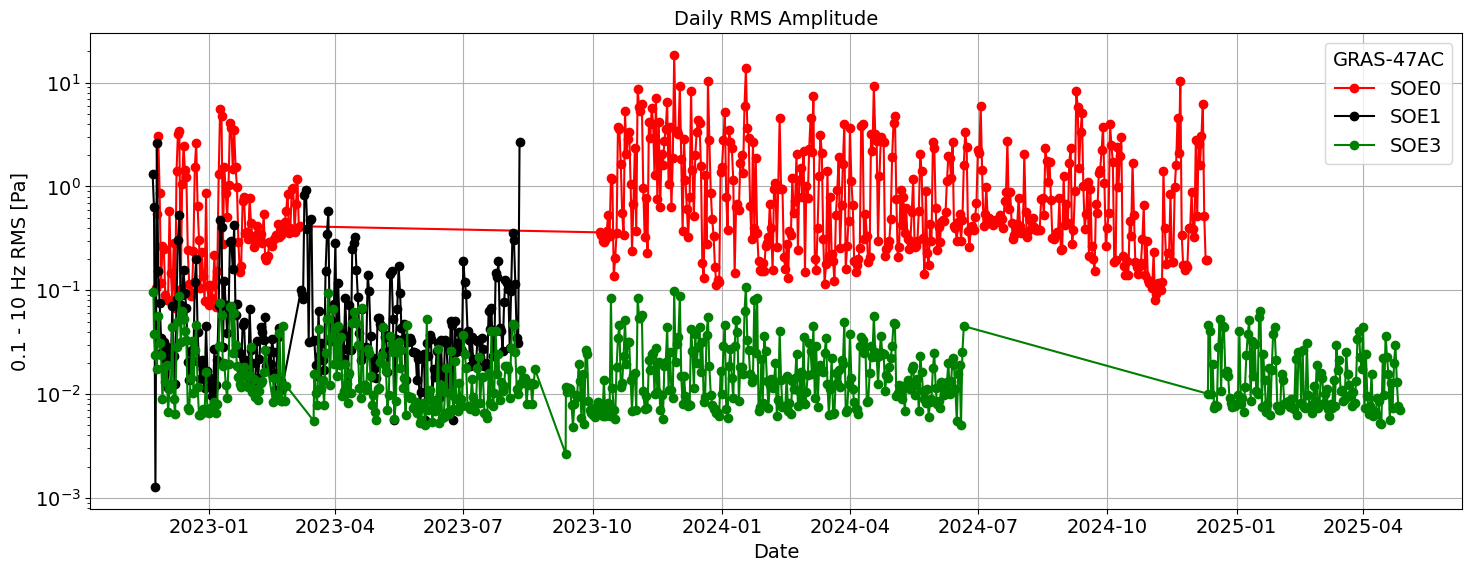}
    \caption{RMS sound pressure in the 0.1--10~\si{\hertz} band for stations SOE0, SOE1 and SOE3, based on 536, 262 and 674 days of valid data, respectively, covering the periods from 24 November 2022 to 11 December 2024 (SOE0), from 22 November 2022 to 10 August 2023 (SOE1) and from 22 November 2022 to 27 April 2025 (SOE3).}
    \label{rms}
\end{figure}

The recording periods varied across the stations. SOE0 collected data from 24 November 2022 to 11 December 2024 (748-day span), of which 536 days contained valid measurements. SOE1 was active from 22 November 2022 until 10 August 2023, a period of 262 days, before its decommissioning. SOE3 acquired the longest dataset, operating from 22 November 2022 until 27 April 2025 (an 887-day deployment), with 674 days of valid measurements. These durations are summarised in Figure~\ref{rms}, which shows the daily RMS sound pressure in the 0.1--10~\si{\hertz} band.

\subsection{Weather Data}

To investigate the influence of atmospheric conditions on the infrasound measurements, meteorological data were obtained from the Weather Underground station IBENET13, located in Benetutti (Giuncana, Sardinia, Italy), approximately \SI{23}{\kilo\metre} west of the Sos Enattos mine. The station is part of the global Weather Underground personal weather station (PWS) network \cite{WU2025}, which provides publicly accessible, near-real-time meteorological observations collected by distributed sensors. The dataset covers the same period as the infrasound measurements, from 22 November 2022 to 27 April 2025. The IBENET13 station records key meteorological parameters, including air temperature, dew point, relative humidity, atmospheric pressure, wind speed, and precipitation, with a temporal resolution of five minutes. In this study, temperature, relative humidity, atmospheric pressure, and wind speed were selected as the primary variables relevant to infrasound generation and propagation \cite{Bowman2005, Fiorucci2018}.

To enable a direct comparison between atmospheric conditions and acoustic measurements, the infrasound data were segmented into five-minute intervals and synchronised with the corresponding meteorological records. For consistency and to avoid temporal bias, one representative five-minute segment per hour (e.g., 12:00--12:05) was selected, resulting in up to 24 samples per day for each station. Each infrasound segment starts at the time stamp of the five-minute weather record with which it is paired, so that both datasets refer to the same five-minute interval and share the same time resolution; no additional temporal averaging was applied to either dataset. This procedure yielded 536, 262, and 674 days of data for stations SOE0, SOE1, and SOE3, respectively, reflecting their respective operational periods. Within each five-minute segment, the ASD of the sound pressure was computed with Welch's method using Hann windows of 4096 samples with 50\% overlap. At the sampling rate of \SI{250}{\hertz} (SOE0 and SOE3), this corresponds to a window length of \SI{16.4}{\second}, a frequency resolution of \SI{0.061}{\hertz} and about 35 averaged windows per segment. At \SI{100}{\hertz} (SOE1), it corresponds to a window length of \SI{41}{\second}, a frequency resolution of \SI{0.024}{\hertz} and about 13 averaged windows.

For the analysis, each meteorological parameter was divided into bins spanning its full range of observed values. Within each bin, the median amplitude spectral density (ASD) of the corresponding infrasound data was calculated. This approach allows for a robust assessment of the dependence of infrasound noise levels on atmospheric conditions while reducing the influence of outliers and short-term fluctuations. Since the meteorological parameters are partly correlated with each other and with sea state, the resulting dependencies are interpreted as associations, and the most relevant correlation, between wind speed and sea state, is examined explicitly in Section~\ref{sec:seasonal}.

\subsection{Marine Activity}

To complement the ground-based meteorological observations, wave reanalysis data were used to evaluate the influence of marine activity on local infrasound levels. Ocean-generated microbaroms are a well-known source of infrasound \cite{Posmentier1967, Waxler2006}, and their intensity is strongly correlated with sea state conditions, particularly wave height \cite{LonguetHiggins1950, Ardhuin2015}. For this purpose, the Mediterranean Sea Waves Reanalysis product (MEDSEA\_MULTIYEAR\_WAV\_006\_012), provided by the Copernicus Marine Environment Monitoring Service (CMEMS) \cite{Korres2021}, was used. This dataset is generated using version 4.6.2 of the third-generation spectral wave model WAM \cite{WAMDI1988}, which simulates the evolution of ocean wave spectra. The model employs nested computational grids to accurately capture swell propagation from the Atlantic Ocean through the Strait of Gibraltar into the Mediterranean Sea. The reanalysis spans the period from January 1985 to the present, with a spatial resolution of $1/24^{\circ}$ ($\sim$\SI{4}{\kilo\metre}) and a temporal resolution of one hour. The dataset incorporates data assimilation from satellite altimetry missions, including Sentinel-1 observations available since 2017, which improves the accuracy of the wave parameters. In this work, the significant wave height (SWH) was selected as the primary variable, serving as a proxy for the strength of marine infrasound sources. Each hourly SWH value was assigned to the five-minute infrasound segment recorded in the same hour. This parameter was subsequently correlated with the infrasound measurements to assess the contribution of ocean wave activity to the observed acoustic field. Since high sea states are frequently associated with strong winds, the SWH dependence was additionally evaluated for low-wind conditions only, in order to separate the marine contribution from that of wind (Section~\ref{sec:seasonal}).

\section{Seasonal Analysis}
\label{sec:seasonal}

To investigate the seasonal variability of the infrasound background at the Sos Enattos site, the dataset was divided into two consecutive periods: \textbf{Part~1}, from 22 November 2022 to 22 September 2023, and \textbf{Part~2}, from 23 September 2023 to 22 September 2024. This division follows the change in instrumental configuration and data availability visible in the RMS time series in Figure~\ref{rms}. Seasons are defined by their astronomical boundaries (spring from 20 March, summer from 21 June, autumn from 23 September and winter from 21 December). For each season, one 5-minute segment per hour with co-located weather data was analysed. In Part~1, SOE0 data are available only for autumn (22 November to 20 December 2022) and winter. SOE1 was operational until 10 August 2023 and is therefore absent from Part~2; for this station, segments with an RMS sound pressure below \SI{1e-3}{\pascal}, corresponding to periods of sensor malfunction, were excluded. In Part~2, the SOE3 summer contains only 17 segments, and a gap in the infrasound recordings between 1 September and 7 October 2023 reduces the number of autumn segments. The number of segments and the 5th, 50th and 95th percentiles of the ASD at \SI{1}{\hertz} for each station, period and season are summarised in Table~\ref{tab:seasonal}.

\begin{figure}
\centering
\includegraphics[width=\textwidth, height=14cm]{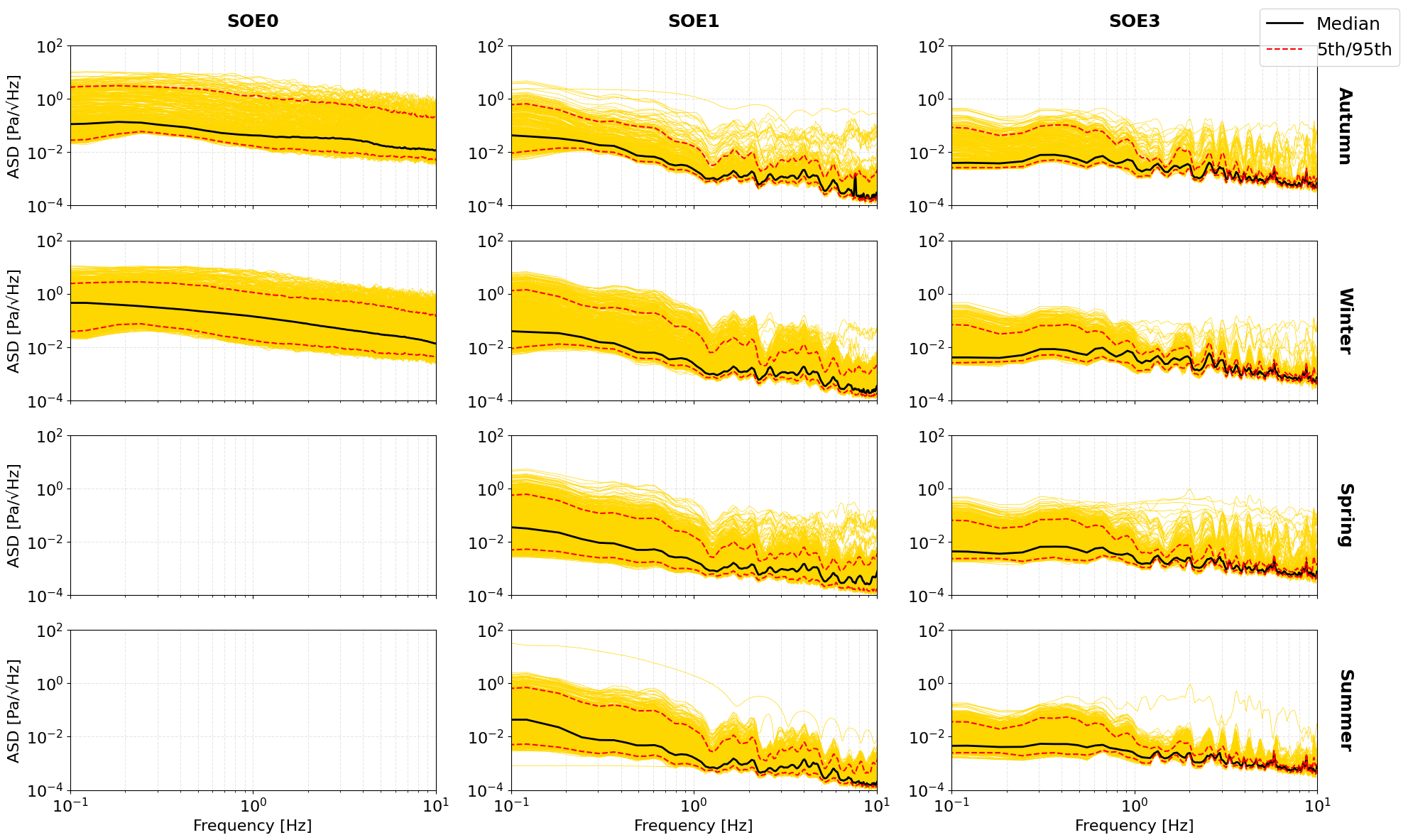}
\caption{ASD of the sound pressure per season for Part~1 (2022-11-22 to 2023-09-22) at SOE0, SOE1 and SOE3. Yellow lines show individual 5-minute segments, the black line the median and the red dashed lines the 5th and 95th percentiles. Empty panels correspond to seasons without data.}
\label{fig:seasonal_part1}
\end{figure}

\begin{figure}
\centering
\includegraphics[width=\textwidth, height=14cm]{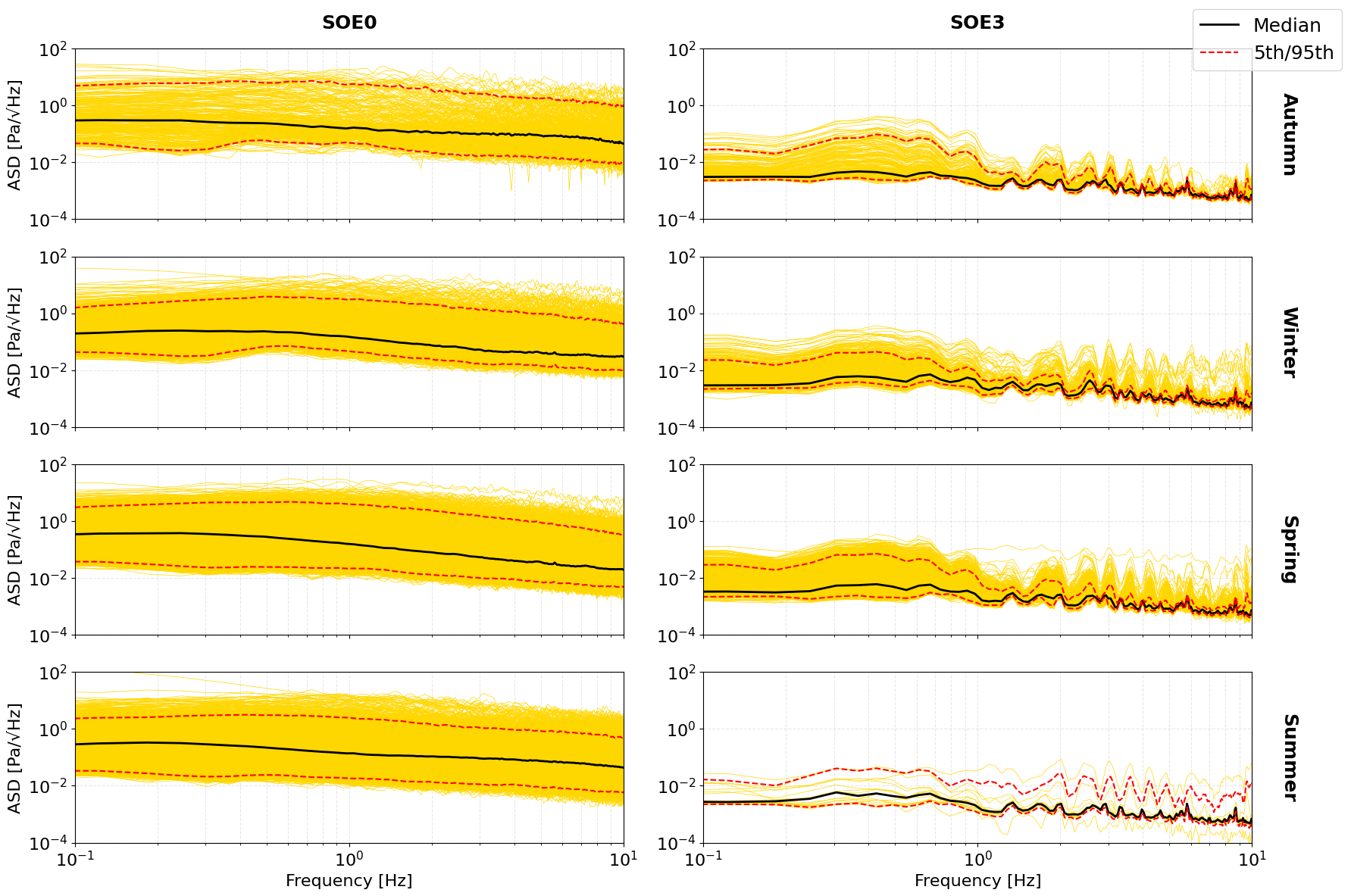}
\caption{As Figure~\ref{fig:seasonal_part1}, for Part~2 (2023-09-23 to 2024-09-22). Only SOE0 and SOE3 are available in this period; the SOE3 summer contains only 17 segments.}
\label{fig:seasonal_part2}
\end{figure}

Figures~\ref{fig:seasonal_part1} and~\ref{fig:seasonal_part2} show the ASD of the sound pressure for each season in Part~1 and Part~2, respectively, together with the median and the 5th and 95th percentiles. In all seasons, the spectra at the underground stations are strongly reduced with respect to the surface over the whole analysed band, starting from \SI{0.1}{\hertz}. At \SI{0.1}{\hertz}, the median ASD at SOE1 and SOE3 is about one and two orders of magnitude lower than at SOE0, respectively. At \SI{1}{\hertz}, the reduction reaches a factor of 25 to 70 (Table~\ref{tab:seasonal}). The spread between the 5th and 95th percentiles is largest at SOE0, where it reaches about two orders of magnitude, and underground it is largest below about \SI{1}{\hertz}. Above \SI{1}{\hertz}, the underground spectra show narrow-band features that are absent at the surface, indicating local sources.

At the surface station SOE0, the median sound pressure shows no consistent seasonal modulation. Over the full seasonal cycle of Part~2, the four seasonal medians at \SI{1}{\hertz} agree within \SI{1.2}{\decibel}. In Part~1, the winter median is \SI{10.5}{\decibel} above the autumn median. However, the Part~1 autumn period covers only the first four weeks after installation and is lower than all other seasons of both periods, while the Part~1 winter median agrees with the Part~2 levels; we therefore do not interpret this difference as a seasonal effect. At SOE0, seasonal differences appear mainly in the tails of the distribution: in Part~2, the 95th percentile at \SI{1}{\hertz} is highest in autumn and spring and lowest in summer.

At the underground stations, in contrast, a winter maximum is observed. At SOE3, the winter median at \SI{1}{\hertz} exceeds the summer median by \SI{7.3}{\decibel} in Part~1 and by \SI{6.9}{\decibel} in Part~2 (\SI{4.4}{\decibel} with respect to spring, given the small number of summer segments in Part~2). At SOE1, the winter median exceeds the summer median by \SI{3.7}{\decibel}. At SOE3 in Part~2, the winter distribution is also narrower than in the other seasons, with a higher 5th and a lower 95th percentile, indicating an elevated background level rather than more frequent high-noise episodes. A possible interpretation is that local wind noise dominates the surface measurements in all seasons, whereas underground, where it is strongly suppressed, the seasonal cycle of large-scale atmospheric infrasound and of oceanic microbaroms, which are stronger during the Northern Hemisphere winter, becomes visible. A seasonal dependence of underground activity, such as ventilation, cannot be excluded.

\begin{figure}
\centering
\includegraphics[width=\textwidth, height=7cm]{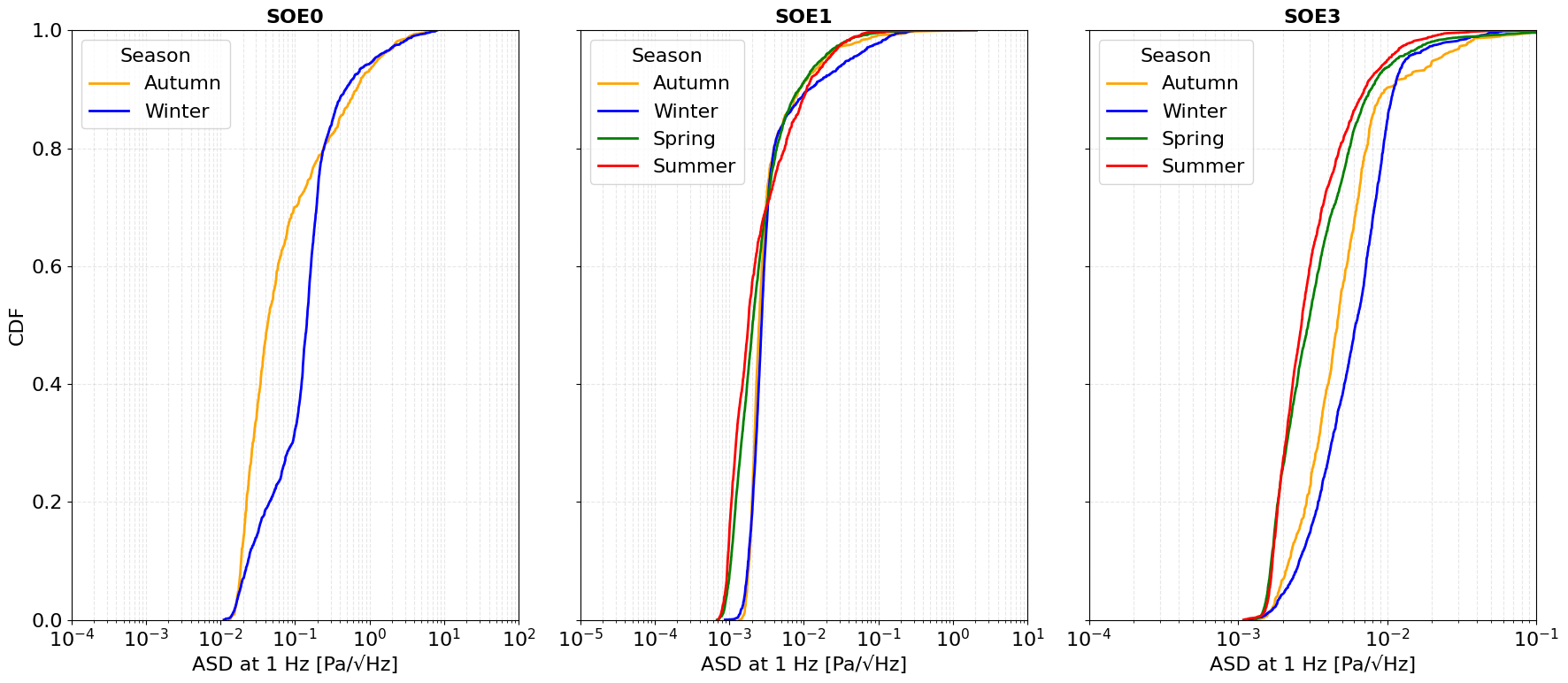}\\
\includegraphics[width=\textwidth, height=7cm]{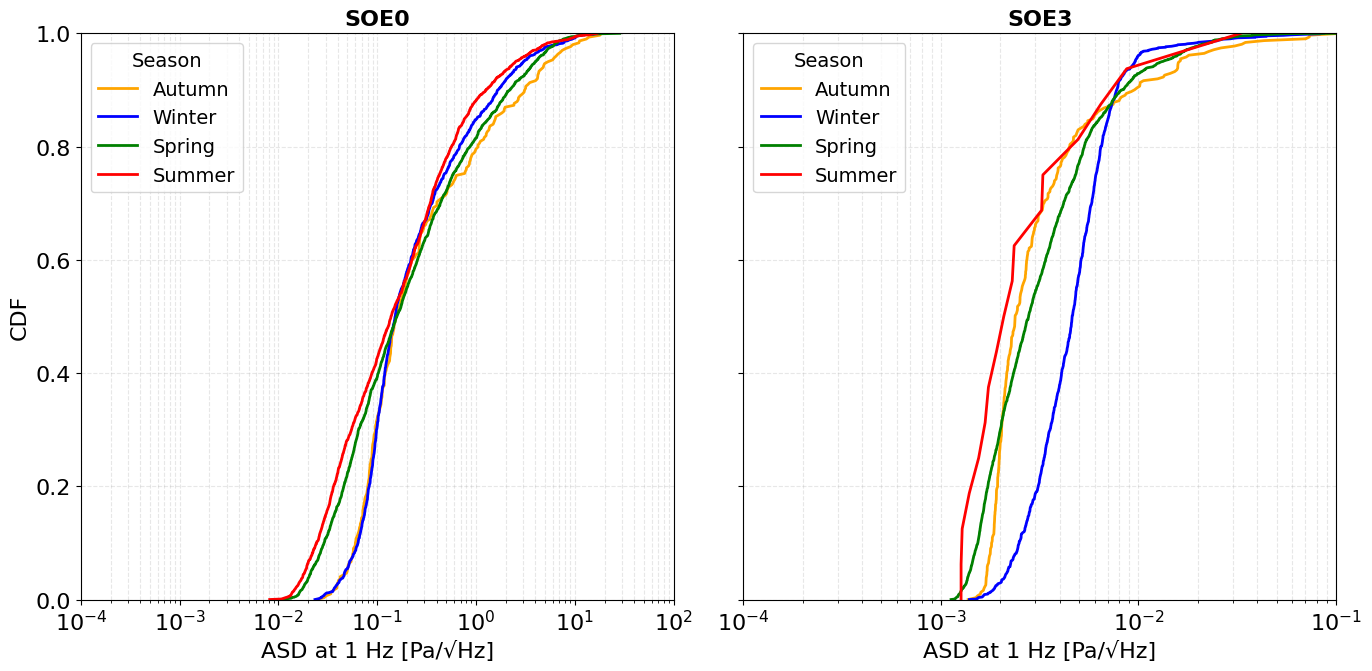}
\caption{Cumulative distribution functions (CDFs) of the ASD at \SI{1}{\hertz} per station and season. Top: Part~1 (2022-11-22 to 2023-09-22). Bottom: Part~2 (2023-09-23 to 2024-09-22). At SOE3, winter is shifted towards higher amplitudes in both periods; at SOE0, the seasonal distributions of Part~2 coincide in their central part and differ mainly in their tails. The step-like SOE3 summer curve in Part~2 reflects the small number of segments (17).}
\label{fig:asd_cdfs}
\end{figure}

The cumulative distribution functions (CDFs) of the ASD at \SI{1}{\hertz} (Figure~\ref{fig:asd_cdfs}) confirm these results. At SOE3, the winter distribution is shifted towards higher amplitudes in both periods, followed by autumn, while spring and summer show the lowest levels. In Part~2, the winter distribution is also the steepest, with the shortest upper tail, and the highest amplitudes occur in autumn. At SOE1, the distributions of the four seasons largely overlap in their central part. Spring and summer are shifted slightly towards lower amplitudes at the low end, and winter shows the most extended upper tail. At SOE0, the Part~1 autumn distribution is shifted towards lower amplitudes for about 80\% of the segments, while the upper tails of autumn and winter are similar. In Part~2, the central parts of the SOE0 distributions coincide for all seasons. The seasons differ at both ends: spring and summer reach lower amplitudes under quiet conditions, and the upper tail is most extended in autumn and least extended in summer.

The comparison between Part~1 and Part~2 at SOE0 indicates an instrumental contribution. While the winter medians of the two periods agree within \SI{0.5}{\decibel}, the 5th percentile at \SI{1}{\hertz} is about 2.7 times higher in Part~2 in both autumn and winter, and the winter 95th percentile is about 2.7 times higher as well. A change in sensor gain or calibration would scale all percentiles by the same factor. The observed pattern, in which the median of the same season is unchanged while the quiet-condition level and the upper tail increase, instead points to a change in the exposure of the sensor to wind and local noise between the two periods. At SOE3, where the installation was unchanged, the seasonal medians of the two periods agree within \SIrange{0.5}{2.5}{\decibel} for winter and spring, and winter is the loudest and summer the quietest season in both periods.

\begin{table}
\centering
\caption{Number of 5-minute segments $N$ and 5th, 50th and 95th percentiles of the ASD at \SI{1}{\hertz} (in \si{\pascal\per\sqrt{\hertz}}) per station, period and season.}
\label{tab:seasonal}
\begin{tabular}{llrccc}
\hline
Station & Season & $N$ & 5th & Median & 95th \\
\hline
\multicolumn{6}{l}{\textit{Part 1 (2022-11-22 to 2023-09-22)}}\\
SOE0 & Autumn & 631  & \num{1.75e-2} & \num{4.26e-2} & \num{1.28} \\
     & Winter & 1828 & \num{1.82e-2} & \num{1.43e-1} & \num{1.13} \\
SOE1 & Autumn & 682  & \num{2.41e-4} & \num{2.40e-3} & \num{1.70e-2} \\
     & Winter & 1831 & \num{1.64e-3} & \num{2.70e-3} & \num{4.10e-2} \\
     & Spring & 2208 & \num{9.38e-4} & \num{2.05e-3} & \num{1.72e-2} \\
     & Summer & 1220 & \num{8.58e-4} & \num{1.78e-3} & \num{2.03e-2} \\
SOE3 & Autumn & 681  & \num{1.88e-3} & \num{4.63e-3} & \num{2.10e-2} \\
     & Winter & 1553 & \num{2.07e-3} & \num{6.13e-3} & \num{1.33e-2} \\
     & Spring & 2208 & \num{1.58e-3} & \num{2.95e-3} & \num{1.14e-2} \\
     & Summer & 1718 & \num{1.62e-3} & \num{2.66e-3} & \num{1.01e-2} \\
\hline
\multicolumn{6}{l}{\textit{Part 2 (2023-09-23 to 2024-09-22)}}\\
SOE0 & Autumn & 300  & \num{4.97e-2} & \num{1.55e-1} & \num{5.38} \\
     & Winter & 1140 & \num{4.73e-2} & \num{1.52e-1} & \num{3.01} \\
     & Spring & 2200 & \num{2.14e-2} & \num{1.59e-1} & \num{4.01} \\
     & Summer & 2256 & \num{1.86e-2} & \num{1.38e-1} & \num{2.51} \\
SOE3 & Autumn & 312  & \num{1.71e-3} & \num{2.38e-3} & \num{1.61e-2} \\
     & Winter & 1140 & \num{2.18e-3} & \num{4.62e-3} & \num{9.59e-3} \\
     & Spring & 2224 & \num{1.41e-3} & \num{2.78e-3} & \num{1.31e-2} \\
     & Summer & 17   & \num{1.27e-3} & \num{2.08e-3} & \num{1.36e-2} \\
\hline
\end{tabular}
\end{table}

\subsection{Influence of Meteorological and Marine Conditions}

\begin{figure}
\centering
\includegraphics[width=\textwidth, height=10cm, keepaspectratio]{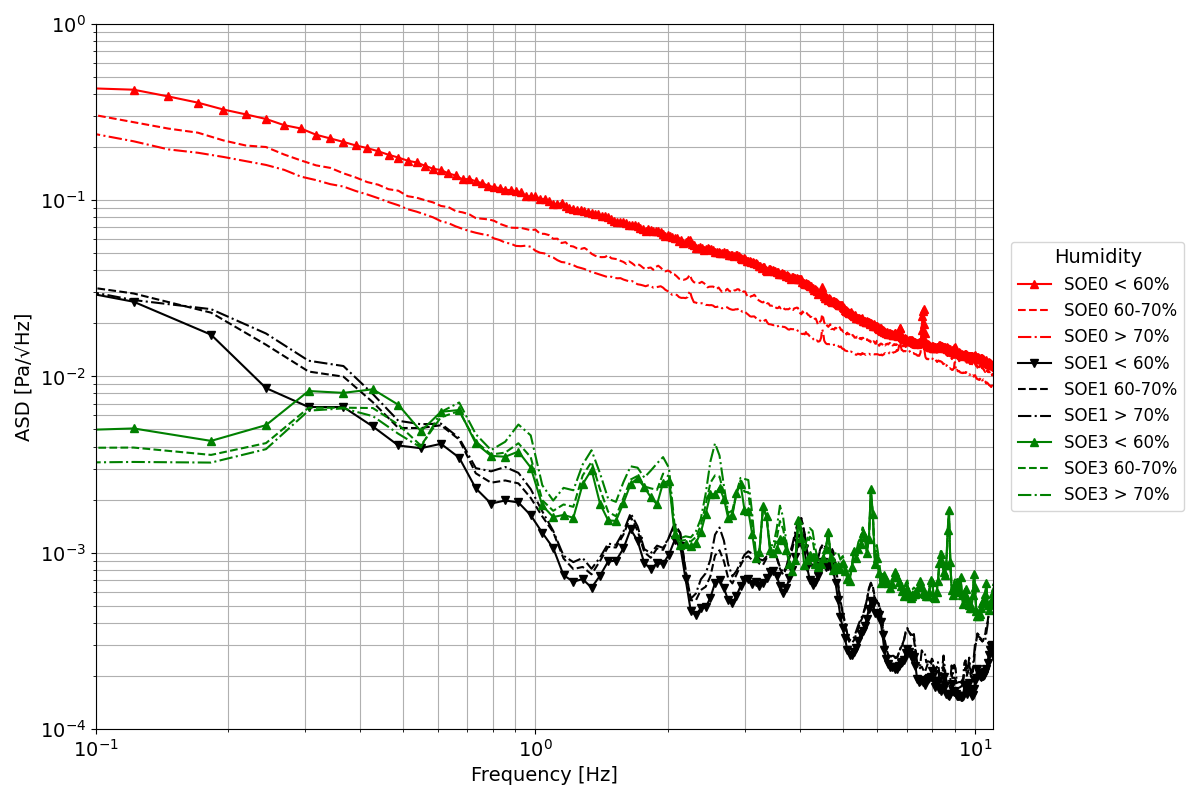}
\caption{Median ASD as a function of relative humidity. At SOE0, the ASD increases with humidity, particularly above 70\%. SOE1 shows a weaker dependence, while SOE3 is largely unaffected.}
\label{fig:hum}
\end{figure}

\begin{figure}
\centering
\includegraphics[width=\textwidth, height=10cm, keepaspectratio]{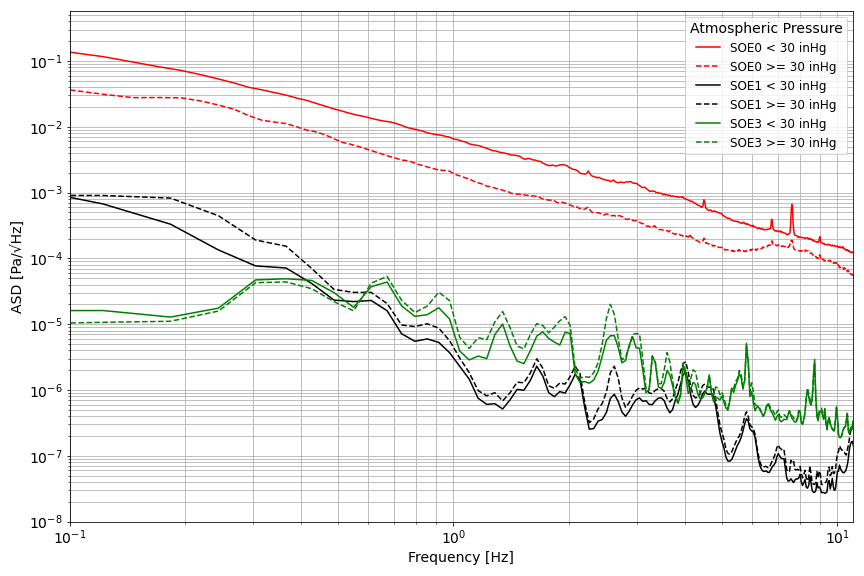}
\caption{Median ASD as a function of atmospheric pressure. SOE0 shows enhanced amplitudes for pressures below \SI{1016}{\hecto\pascal}, consistent with storm activity, whereas SOE1 and SOE3 show only minor variations.}
\label{fig:p}
\end{figure}

\begin{figure}
\centering
\includegraphics[width=\textwidth, height=18cm, keepaspectratio]{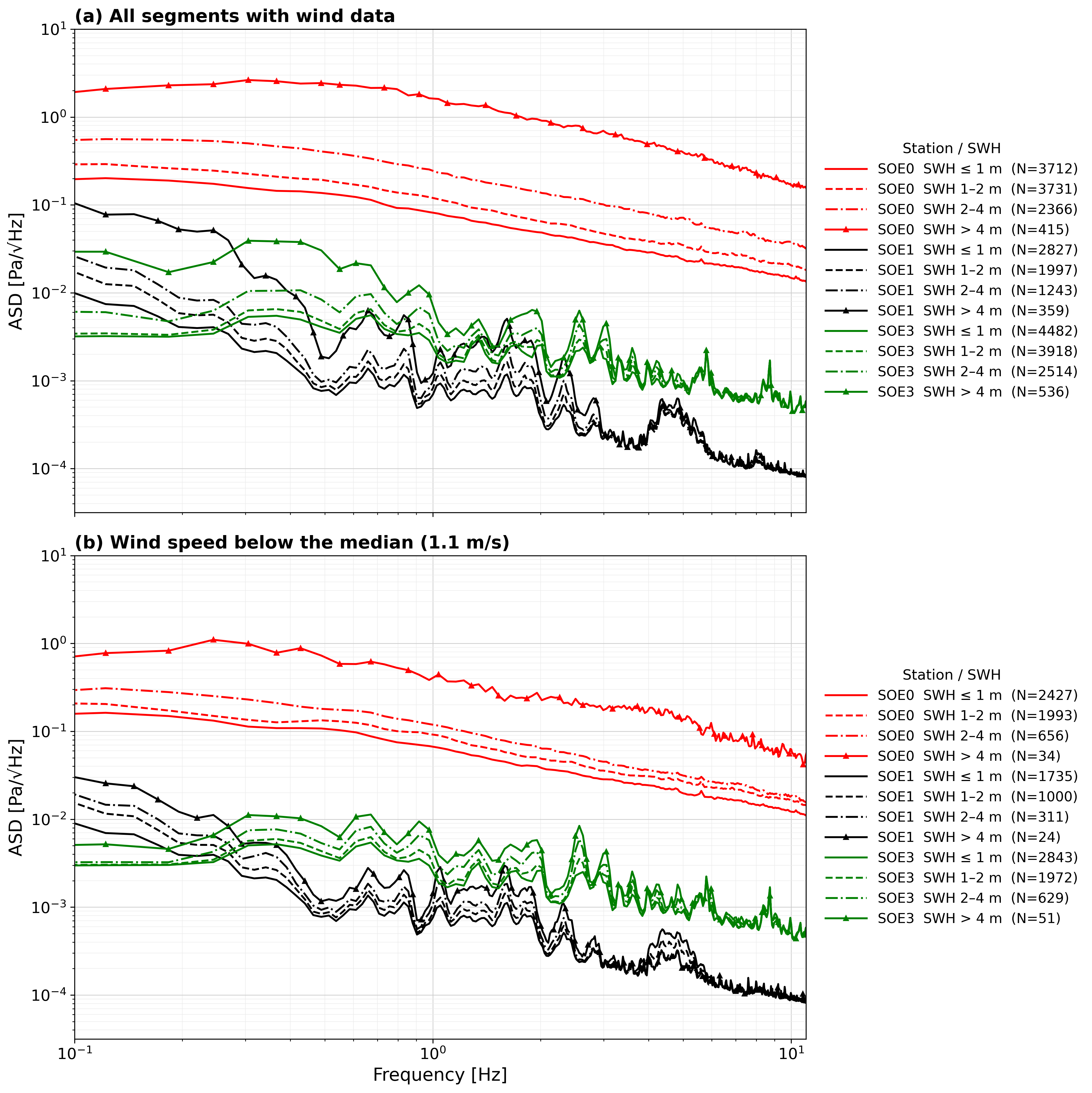}
\caption{Median ASD for each significant-wave-height (SWH) class at SOE0 (red), SOE1 (black) and SOE3 (green). (a) All segments with co-located wind data. (b) Only segments with wind speed below the median of the distribution (\SI{1.1}{\metre\per\second}). The number of segments in each class is given in the legends.}
\label{fig:swh_lowwind}
\end{figure}

\begin{figure}
\centering
\includegraphics[width=\textwidth, height=9cm, keepaspectratio]{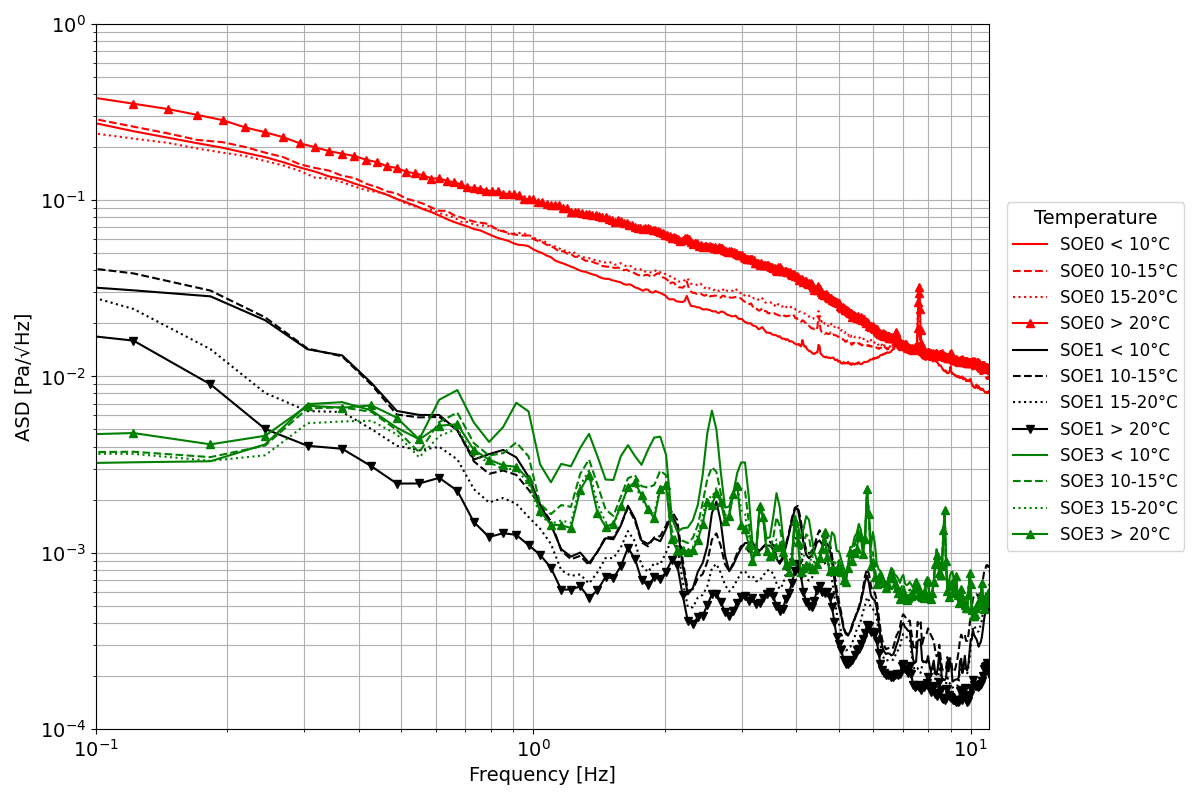}
\caption{Median ASD as a function of temperature. SOE0 shows higher amplitudes above \SI{20}{\celsius} and lower amplitudes below \SI{10}{\celsius}. SOE1 shows a weaker dependence, and SOE3 is largely insensitive.}
\label{fig:temp}
\end{figure}

\begin{figure}
\centering
\includegraphics[width=\textwidth, height=9cm, keepaspectratio]{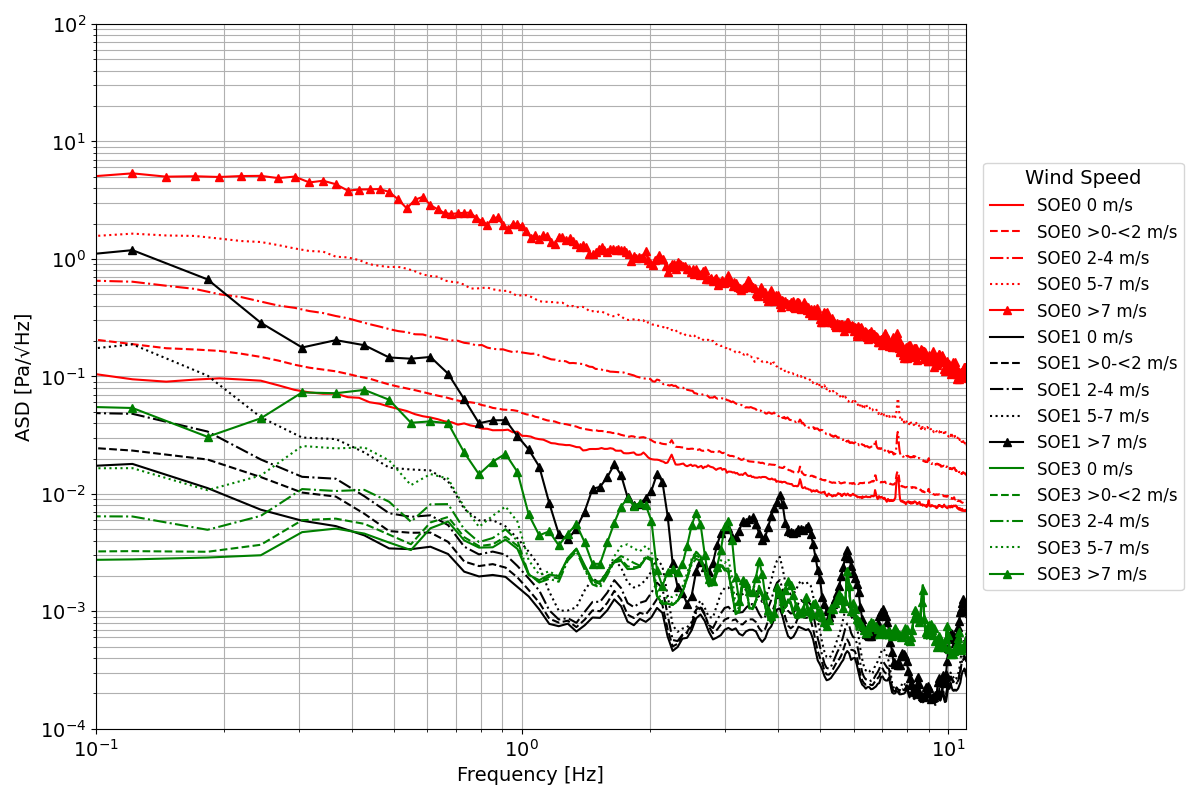}
\caption{Median ASD as a function of wind speed. Wind is the dominant factor influencing the infrasound level. SOE0 shows the strongest increase with wind speed, by more than an order of magnitude between calm conditions and wind speeds above \SI{7}{\metre\per\second}. At SOE1 and SOE3, the dependence is weaker and concentrated below about \SI{1}{\hertz}.}
\label{fig:wind}
\end{figure}

To quantify the impact of environmental drivers on infrasound, the matched infrasound--weather dataset was analysed with respect to relative humidity, atmospheric pressure, significant wave height (SWH), temperature and wind speed (Figures~\ref{fig:hum}, \ref{fig:p}, \ref{fig:swh_lowwind}, \ref{fig:temp} and~\ref{fig:wind}). Median ASD curves were computed in bins of each parameter for the three stations. Since these parameters are not independent of each other (for example, strong winds, high sea states, low pressure and season are often linked through the same weather systems), the dependencies shown in the individual figures should be read as associations rather than independent effects. For the most evident case, sea state and wind, the two contributions were separated explicitly, as discussed below.

Relative humidity mainly affects the surface station SOE0, where infrasound amplitudes increase with humidity and are highest above 70\% in the \SIrange{0.1}{10}{\hertz} range. At SOE1 this dependence is weaker, and at SOE3 it is negligible. Atmospheric pressure has its strongest influence at SOE0, where pressures below \SI{1016}{\hecto\pascal} are associated with elevated infrasound levels, consistent with the passage of synoptic-scale disturbances and the associated strong winds. At SOE1 and SOE3, the effect of pressure is minimal. Temperature introduces only modest variations: at SOE0, higher infrasound levels are observed above \SI{20}{\celsius} and lower levels below \SI{10}{\celsius}, consistent with enhanced daytime convective turbulence. SOE1 shows a similar but weaker dependence, and SOE3 is largely insensitive.

Wind speed is the dominant factor at all stations, in agreement with the well-established role of wind as a primary source of infrasonic noise \cite{Bowman2005, Walker2010}. At SOE0, the ASD increases by more than an order of magnitude between calm conditions and wind speeds above \SI{7}{\metre\per\second} over the whole band. At SOE1 and SOE3, the dependence on wind is weaker and concentrated below about \SI{1}{\hertz}, although it remains clearly detectable.

Sea state can contribute through oceanic microbaroms \cite{Posmentier1967, Waxler2006, Ardhuin2015}. At SOE0, the median ASD for SWH above \SI{4}{\metre} is about 16 times higher than for SWH below \SI{1}{\metre} (Figure~\ref{fig:swh_lowwind}a). At the underground stations, a similar increase (about a factor of 8) is found only between \SI{0.1}{\hertz} and \SI{0.5}{\hertz}, while no dependence on SWH is observed between \SI{1}{\hertz} and \SI{10}{\hertz}. However, sea state and wind speed are not independent, since strong winds and high waves are often produced by the same weather systems. To separate the two effects, the SWH analysis was repeated using only segments with wind speed below the median of the distribution (\SI{1.1}{\metre\per\second}; Figure~\ref{fig:swh_lowwind}b). High sea states occur predominantly together with strong winds: only 8\%, 7\% and 10\% of the segments with SWH above \SI{4}{\metre} remain in the low-wind subset at SOE0, SOE1 and SOE3, respectively. Under low-wind conditions, the dependence on SWH is strongly reduced. At the surface, the ratio between the median ASD for SWH above \SI{4}{\metre} and below \SI{1}{\metre} decreases from 16.8 to 7.2 between \SI{0.1}{\hertz} and \SI{0.5}{\hertz}, and from 15.8 to 5.9 between \SI{1}{\hertz} and \SI{10}{\hertz}; for SWH of 2--4\,m, the corresponding ratios decrease from 3.1 to 1.9 and from 2.6 to 1.5. Underground, the ratio for SWH above \SI{4}{\metre} decreases between \SI{0.1}{\hertz} and \SI{0.5}{\hertz} from 8.1 to 2.5 at SOE1 and from 7.3 to 2.1 at SOE3. The residual low-frequency dependence underground is consistent with a contribution from oceanic microbaroms \cite{LonguetHiggins1950, Ardhuin2015}. The broadband residual at the surface is more difficult to interpret: it is based on a small number of segments for the highest sea states (34 for SWH above \SI{4}{\metre}) and may partly reflect wind conditions that are not captured by the single local weather station.

Several features of the data point to anthropogenic contributions, in particular underground. First, the underground spectra show narrow-band features between \SI{1}{\hertz} and \SI{10}{\hertz} that do not depend on wind or sea state and are absent at the surface (Figures~\ref{fig:seasonal_part1}, \ref{fig:seasonal_part2} and~\ref{fig:swh_lowwind}). Second, the deeper station SOE3 shows higher sound-pressure levels than SOE1 above about \SI{1}{\hertz}, which cannot be explained by the attenuation of the surface field with depth. Third, the SOE3 spectra contain narrow lines at about \SI{43}{\hertz} and at its second harmonic at about \SI{86}{\hertz}, a signature typical of rotating electromechanical equipment such as ventilation fans or pumps. The Sos Enattos mine hosts ventilation, pumping and other infrastructure, so these sources are expected to dominate the residual acoustic field underground at frequencies where atmospheric infrasound is strongly attenuated. Their impact on the NN estimate is discussed in Section~\ref{sec:nn}.

Overall, these results show that local wind forcing is the primary driver of infrasound variability at the Sos Enattos site. Most of the apparent dependence on sea state is a consequence of its correlation with wind, with a residual microbarom contribution below \SI{0.5}{\hertz}. Humidity, pressure and temperature play a secondary role, largely confined to the surface station, while local anthropogenic sources shape the underground spectra above about \SI{1}{\hertz}.

\section{Newtonian Noise Contribution Estimation}
\label{sec:nn}

The atmospheric NN contribution is estimated by converting the acoustic pressure fluctuations measured at each station into air-density perturbations and integrating the resulting gravitational acceleration acting on a test mass. Under the linear-acoustic approximation \cite{Saulson1984, Creighton2008, Harms2015, Fiorucci2018}, the density perturbation is
\begin{equation}
\delta\rho(\mathbf{r},t)=\frac{\delta p(\mathbf{r},t)}{v_s^2},
\end{equation}
where $\delta p(\mathbf{r},t)$ is the acoustic pressure fluctuation and $v_s=\SI{330}{\metre\per\second}$ is the speed of sound in air. The gravitational acceleration induced at the test-mass position $\mathbf{r}_t$ follows from Newton's law of gravitation,
\begin{equation}
\mathbf{a}(\mathbf{r}_t,t)=\frac{G}{v_s^2}\int_{V}
\delta p(\mathbf{r},t)\,\frac{\mathbf{r}-\mathbf{r}_t}{\lVert\mathbf{r}-\mathbf{r}_t\rVert^{3}}\,\mathrm{d}V ,
\end{equation}
where the integration extends over the air-filled volume $V$ surrounding the test mass. We adopt \rev{a cavern with the dimensions foreseen in the current ET design, $\SI{175}{\metre}\times\SI{20}{\metre}$ in the horizontal plane and \SI{30}{\metre} in height, with its long axis along the arm}, from which the evacuated vacuum tower (\rev{\SI{20}{\metre}} tall, \SI{1}{\metre} radius) is excluded, as it contains negligible air mass. We use a right-handed Cartesian reference system with the origin on the tower axis at floor level, the $x$ axis along the interferometer arm, the $y$ axis horizontal and perpendicular to the arm, and the $z$ axis pointing vertically upwards. The test mass is located on the tower axis at $\mathbf{r}_t=(0,0,\SI{2}{\metre})$. \rev{Two tower positions are considered: at the centre of the cavern and \SI{10}{\metre} from its end wall.} The volume is discretised into \rev{cubic voxels of \SI{0.2}{\metre} within \SI{10}{\metre} of the tower axis and \SI{0.5}{\metre} elsewhere}. The same geometry is used for \rev{both underground stations}, so that differences between \rev{their} estimates reflect only differences in the locally measured pressure field.

The acoustic field is modelled as an isotropic superposition of plane waves, $\delta p\propto e^{\mathrm{i}k\,\hat{\mathbf{n}}\cdot\mathbf{r}}$ with wavenumber $k=2\pi f/v_s$, arriving from all directions with independent random phases \cite{Creighton2008, Fiorucci2018}. Only the acceleration component along the arm direction $\hat{\mathbf{x}}$ couples to the interferometer signal. For a propagation direction $\hat{\mathbf{n}}$, the corresponding geometrical kernel is evaluated as the discrete sum
\begin{equation}
K(\hat{\mathbf{n}},f)=\sum_{j} \Delta V_j\,e^{\mathrm{i}k\,\hat{\mathbf{n}}\cdot(\mathbf{r}_j-\mathbf{r}_t)}\,
\frac{x_j-x_t}{\lVert\mathbf{r}_j-\mathbf{r}_t\rVert^{3}},
\end{equation}
where the sum runs over all air voxels $\mathbf{r}_j$ \rev{of volume $\Delta V_j$}. Because waves from different directions carry independent phases, their contributions add in power. Averaging over $N$ directions distributed uniformly over the full sphere gives the transfer function from pressure to acceleration amplitude spectral density,
\begin{equation}
T(f)=\frac{G}{v_s^2}
\left[\frac{1}{N}\sum_{n=1}^{N}\bigl|K(\hat{\mathbf{n}}_n,f)\bigr|^{2}\right]^{1/2},
\end{equation}
such that $S_a^{1/2}(f)=T(f)\,S_p^{1/2}(f)$, where $S_p^{1/2}(f)$ is the measured pressure amplitude spectral density. The directions are placed on a Fibonacci lattice, which samples the sphere nearly uniformly for any $N$. We use $N=256$ and verified that the result does not depend on this choice: reducing the number of directions to $N=64$ changes $T(f)$ by less than \rev{0.6\%, and the voxel grid differs by less than 1.5\% from a uniform \SI{0.1}{\metre} grid}. This fast convergence reflects the smooth dependence of the kernel on the propagation direction, since the acoustic wavelength ($\geq\SI{2.2}{\metre}$ below \SI{150}{\hertz}) is large compared with the voxel size. $T(f)$ increases from \rev{$\SI{1.3e-14}{\metre\per\second\squared\per\pascal}$ at \SI{1}{\hertz} to $\SI{2.4e-14}{\metre\per\second\squared\per\pascal}$ at \SI{5}{\hertz} for the test mass at the centre of the cavern, and decreases at higher frequencies, where the contributions of distant air cancel}.

\rev{The pressure measured at the surface station SOE0 acts on an underground test mass through the atmosphere above the ground rather than through the cavern air. For this station, the integration extends over the air half-space above a test mass at a depth $D=\SI{200}{\metre}$. For the same isotropic field, the integral can be evaluated analytically:}
\begin{equation}
T_{\rm s}(f)=\frac{2\pi G}{v_s^2\,k}\left\langle \cos^2\phi\;e^{-2kD\cos\epsilon}\right\rangle^{1/2},
\label{eq:Ts}
\end{equation}
\rev{where $\epsilon$ and $\phi$ are the elevation and the azimuth (with respect to the arm) of the propagation direction, and the average is taken over all directions. A numerical integration over a finite air layer of $\SI{1000}{\metre}\times\SI{1000}{\metre}\times\SI{300}{\metre}$ gives the same result within 15\%.}

The strain-equivalent amplitude spectral density is obtained by converting acceleration to displacement and normalising to the arm length $L=\SI{10}{\kilo\metre}$ of the Einstein Telescope. Assuming uncorrelated NN at the four test masses of an interferometer \cite{Harms2015},
\begin{equation}
S_h^{1/2}(f)=\frac{2\,T(f)\,S_p^{1/2}(f)}{(2\pi f)^{2}\,L}.
\label{eq:strain_conversion}
\end{equation}

Eq.~\eqref{eq:strain_conversion} is applied to the sound-pressure spectra measured \emph{in situ} at the surface station SOE0 \rev{(with $T_{\rm s}$)} and at the underground stations SOE1 (\SI{84}{\metre} depth) and SOE3 (\SI{160}{\metre} depth). For each station, the pressure ASD is computed for one 5-minute segment per hour, and the 5th, 50th and 95th percentiles are taken in each frequency bin, based on 12\,786 segments for SOE0, 5\,950 for SOE1 and 15\,406 for SOE3. For SOE1, segments with an RMS sound pressure below \SI{1e-3}{\pascal}, corresponding to periods of sensor malfunction, were excluded. For the underground stations, the attenuation of the surface infrasound field is therefore not modelled but directly captured by the local measurement: the NN estimate reflects the residual acoustic field actually present underground. Spectra are shown up to $0.8$ times the Nyquist frequency of each recording (\SI{100}{\hertz} for SOE0 and SOE3, \SI{40}{\hertz} for SOE1), above which the anti-aliasing filter of the digitiser attenuates the signal.

\begin{figure}
    \centering
    \includegraphics[width=0.95\textwidth]{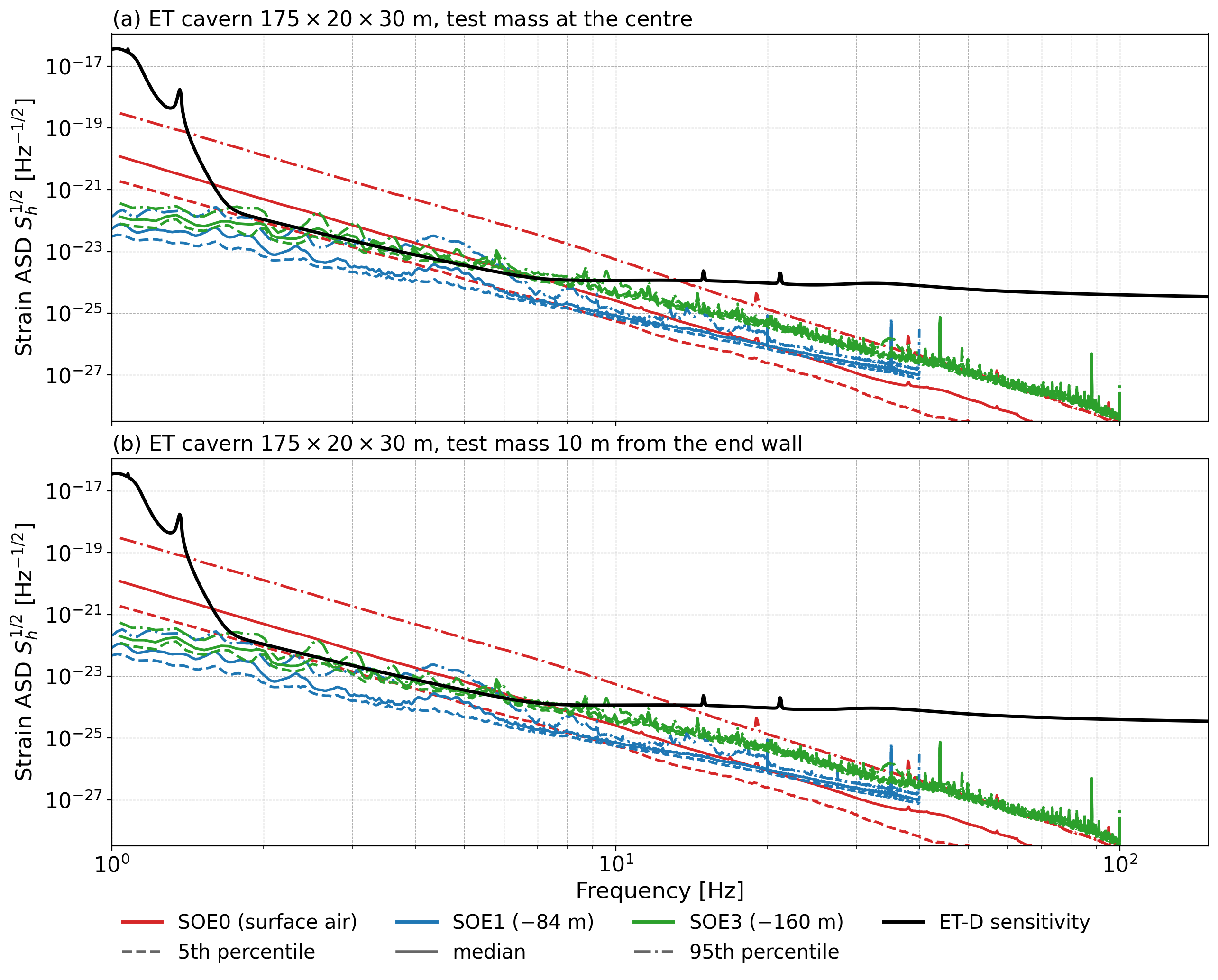}
    \caption{\rev{NN strain ASD from atmospheric infrasound compared with the
ET-D sensitivity (black). (a) Test mass at the centre of an ET cavern of
$175\times20\times30$~m$^3$; (b) test mass 10~m from the end wall of the same cavern. Blue
and green: noise from the pressure measured at SOE1 ($-84$~m) and SOE3 ($-160$~m),
integrated over the cavern air. Red: noise from the pressure measured at the surface
(SOE0), integrated over the atmosphere above a test mass 200~m underground. Dashed,
solid and dash-dotted lines are the 5th percentile, median and 95th percentile.}}
    \label{fig:nn_characteristic}
\end{figure}

Figure~\ref{fig:nn_characteristic} shows that atmospheric infrasound \rev{at the surface remains} a significant noise source \rev{even for a test mass \SI{200}{\metre} underground}. The median SOE0 NN estimate exceeds the ET-D sensitivity \cite{Hild2011} between \rev{\SI{1.6}{\hertz} and \SI{7.1}{\hertz}, by up to a factor of 4.7}, and the 95th percentile between \rev{\SI{1.5}{\hertz} and \SI{13}{\hertz}, by up to a factor of 130}. \rev{Under the quietest conditions (5th percentile), it remains just below the design sensitivity.}

\rev{Above about \SI{3}{\hertz}, the NN is dominated by the acoustic field inside the cavern. At SOE1 the median lies a factor of 1.4 below the ET-D curve, while the 95th percentile exceeds it between \SI{3.1}{\hertz} and \SI{6.1}{\hertz}. At SOE3 the median exceeds ET-D between \SI{2.5}{\hertz} and \SI{8.7}{\hertz}, by up to a factor of 3.5, and the 95th percentile by up to a factor of 4.9; with the test mass \SI{10}{\metre} from the end wall these factors become 2.6 and 3.6.} Although deeper, SOE3 shows NN levels comparable to or up to a few times higher than SOE1 above \SI{3}{\hertz}. This indicates that, below the surface, the residual acoustic field is determined by local, anthropogenic sources rather than by depth alone (Section~\ref{sec:seasonal}). The narrow-band features at about \SI{43}{\hertz} and \SI{86}{\hertz} at SOE3, attributed to rotating electromechanical equipment, remain well below the ET-D sensitivity. We note that SOE1 was operational only between November 2022 and August 2023, so its percentiles do not cover a full seasonal cycle.

The steep decrease of all NN curves with frequency reflects the $(2\pi f)^{-2}$ acceleration-to-strain conversion in Eq.~\eqref{eq:strain_conversion}, partly compensated by the increase of $T(f)$ up to about \rev{\SI{5}{\hertz}}. Atmospheric NN is therefore mainly relevant for the low-frequency part of the detector noise budget. Overall, these results confirm the clear advantage of underground placement for next-generation gravitational-wave observatories \cite{Beker2011, Brundu2022}. At the same time, the \rev{excess} at SOE3 shows that controlling local acoustic sources in the underground halls will be as important as depth in keeping the infrasound NN below the design sensitivity.


\section{Conclusions}
\label{sec:conclusions}

In this work, we presented a long-term analysis of infrasound measurements at the Sos Enattos site, using GRAS 47AC microphones deployed at the surface (SOE0) and at depths of \SI{84}{\metre} (SOE1) and \SI{160}{\metre} (SOE3). The study combined spectral, statistical and environmental analyses to characterise the infrasound field and to assess its contribution to atmospheric Newtonian noise (NN) for the Einstein Telescope (ET).

At the surface, the median infrasound level shows no consistent seasonal modulation: over a full year of data, the seasonal medians at \SI{1}{\hertz} agree within about \SI{1}{\decibel}, and seasonal differences appear mainly in the tails of the distribution. Underground, where local wind noise is strongly suppressed, a winter maximum is observed at both stations, reaching about \SI{7}{\decibel} above summer at SOE3 in both years of data and about \SI{4}{\decibel} at SOE1. Wind speed is the dominant environmental driver of infrasound variability. The apparent dependence on sea state is largely a consequence of the correlation between sea state and wind: when only low-wind periods are considered, the dependence on significant wave height is strongly reduced, and at the underground stations it is confined to frequencies below \SI{0.5}{\hertz}, consistent with a residual contribution from oceanic microbaroms. Relative humidity and temperature play a secondary role, with effects largely confined to the surface station.

The infrasound field is strongly attenuated underground, starting from \SI{0.1}{\hertz}: the median sound-pressure ASD at SOE1 and SOE3 is reduced by one to two orders of magnitude with respect to the surface. The reduction, however, is not monotonic with depth: above about \SI{1}{\hertz}, SOE3 shows sound-pressure levels comparable to or higher than those at the shallower SOE1, together with narrow-band features typical of electromechanical equipment. This indicates that, once underground, the residual acoustic field is governed by local anthropogenic sources rather than by depth alone.

Using the measured sound-pressure spectra, the atmospheric NN was estimated for \rev{an ET-sized cavern} within a linear-acoustic framework with an isotropic plane-wave field. \rev{The surface pressure field, acting through the atmosphere above a test mass at \SI{200}{\metre} depth, exceeds the ET-D sensitivity in median conditions between \SI{1.6}{\hertz} and \SI{7.1}{\hertz}, by up to a factor of 4.7. Inside the cavern, the median estimate at SOE1 remains a factor of 1.4 below ET-D, while at SOE3 it exceeds ET-D between \SI{2.5}{\hertz} and \SI{8.7}{\hertz}, by up to a factor of 3.5. These estimates assume a plane-wave field coherent over the whole cavern.}

Overall, the results show that underground deployment is essential to keep infrasound NN below the ET design sensitivity at Sos Enattos, \rev{but that depth alone is not sufficient: at SOE3 the acoustic field inside an ET-sized cavern already exceeds the design sensitivity between about \SI{2}{\hertz} and \SI{9}{\hertz}.} The measurements reflect the current state of the site, where local sources such as existing mine infrastructure already raise the acoustic level at the deepest station (e.g.\ the narrow-band features at about \SI{43}{\hertz} and \SI{86}{\hertz}). The infrastructure and equipment of a future ET laboratory, such as ventilation and air-conditioning systems, cryogenic plants, vacuum pumps and other machinery in the underground halls and tunnels, may significantly increase the acoustic contribution to the NN budget. The acoustic emission of this equipment will therefore need to be carefully evaluated at the design stage, and effective mitigation measures, such as low-noise equipment, acoustic isolation of machinery and cavern design, will need to be implemented. Together with continued environmental monitoring, this will be required to ensure that atmospheric NN does not limit the low-frequency sensitivity of ET. With these precautions, the infrasound environment of Sos Enattos is compatible with the requirements of a next-generation gravitational-wave detector.

\ifanon\else
\ack{The authors acknowledge support from the National Science Centre, Poland (NCN), grant No.\ 2023/49/B/ST9/02777. TB and DR were supported by the European Union's Horizon Europe research and innovation program under Grant Agreement No.\ 101079696 (HORIZON-INFRA-2021-DEV-02 (ET-PP)) and by the Polish Ministry of Science and Higher Education Project (No.\ W55/HE/2022). This work was also supported by the FuSe COST Action CA24101, and by the Hungarian National Research, Development and Innovation Office (NKFIH) under Contracts NKFIH NKKP-Advanced 150038 and 2025-1.1.5-NEMZ\_KI-2025-00011.}
\fi

\ifanon \hypersetup{pdfauthor={}, pdfcreator={}} \fi
\section*{Data availability statement}
The data that support the findings of this study are available from the
corresponding author upon reasonable request.




\bibliographystyle{iopart-num}
\bibliography{references}

\end{document}